\documentclass{ws-ijmpd}

\usepackage[caption=false]{subfig}
\usepackage[super,compress]{cite}

\begin{document}

\markboth{W. S. Hip\'olito--Ricaldi and J. R. Villanueva}
{A Jacobian generalization of the pseudo Nambu-Goldstone boson potential}

%%%%%%%%%%%%%%%%%%%%% Publisher's Area please ignore %%%%%%%%%%%%%%%
%
\catchline{}{}{}{}{}
%
%%%%%%%%%%%%%%%%%%%%%%%%%%%%%%%%%%%%%%%%%%%%%%%%%%%%%%%%%%%%%%%%%%%%

\title{A Jacobian generalization of the pseudo Nambu--Goldstone boson potential}

\author{W. S. Hip\'olito--Ricaldi}

\address{Department of Physics, McGill University, \\ Montr\'eal, QC, H3A 2T8, Canada\footnote{
Permanent Institution: Departamento de Ci\^encias Naturais, Universidade Federal do Esp\'{\i}rito Santo, ES, Brazil}\\
ricaldiw@physics.mcgill.ca}

\author{J. R. Villanueva}

\address{Instituto de F\'{\i}sica y Astronom\'ia,\\ 
Facultad de Ciencias, Universidad de Valpara\'iso, \\
Gran Breta\~na 1111, Playa Ancha\\
Valpara\'iso, Chile \\
jose.villanuevalob@uv.cl}

\maketitle

\begin{history}
\received{Day Month Year}
\revised{Day Month Year}
\end{history}

\begin{abstract}
We enlarge the classes of inflaton and  quintessence fields by generalizing the pseudo Nambu--Goldstone boson potential
by means of elliptic Jacobian functions, which are characterized by a parameter $k$. 
We use such a generalization to implement an inflationary era and a late acceleration of the universe.
As an inflationary model the Jacobian generalization leads us to a number of e-foldings and a 
primordial spectrum of  perturbations compatible with the {\it Planck collaboration} 2015. As a quintessence 
model, a study of the evolution of its equation of state (EoS) and its $w'$--$w$ plane helps us to 
classify  it  as a thawing  model. This allows us to consider analytical approximations 
for the EoS recently discovered for thawing quintessence. By using JLA Supernovae Ia  and 
Hubble parameter $H(z)$ data sets, we perform an observational analysis of the viability of the model
as quintessence.
\end{abstract}


\ccode{PACS numbers:}

%\tableofcontents

\section{Introduction}
\label{intro}
Scalar fields are widely used to endeavor to explain  some aspects of the
modern cosmology,  in particular, the inflationary epoch in  early 
times and the recent acceleration of the universe. Mechanisms 
for inflation and recent acceleration are not so different. The main
difference is that non-relativistic matter cannot be ignored in recent 
times. Moreover, the energy scale of the quintessence potentials needs 
to be much smaller than the inflationary ones. 

Inflation is a  period preceding the standard cosmology 
where the evolution of the universe can be described by a scalar field, the inflaton.
It was proposed to solve  the horizon and the flatness problems of the standard model,
as well as provide the mechanism to generate the seeds  for the formation of structures  
and anisotropies of the cosmic microwave background (CMB). During the inflationary 
period,  the universe accelerates its expansion to an almost exponential rate,  cooling  and smoothing 
the spatial structure. Furthermore, inflation predicts fluctuations with an almost scale invariant spectrum  which has 
been  confirmed by  recent CMB data \cite{planck,planck15}. To connect with the standard cosmological periods dominated 
by radiation and  matter, it is necessary to add a period of reheating. The usual 
mechanism can also be preceded by a more efficient phase 
known as preheating \cite{kofman,kofman2,lobito1,lobito2}.

On the other hand,  several cosmological experiments have demostrated the existence of a phase of accelerated expansion 
in current times \cite{planck,planck15,late,late2,late3}. This leads us to  one  of the most challenging problems in cosmology: 
the nature of  dark energy.  So far, the most sucessful dark energy candidate 
is the cosmological constant  $\Lambda$. However, this model apparently has shortcomings such as discrepancy between the 
value of the vacuum energy obtained through observations and its  estimated theoretical 
value, or even the coincidence problem \cite{problems,problems2,problems3}. Thus, in recent years, alternatives to solve these 
problems have been sought. Alternatives to the $\Lambda$CDM model can be divided into two 
groups. In the first group, general relativity is modified  to obtain late 
acceleration without exotic components (see for example Refs.~\refcite{modifies}-\refcite{modifies3}).  The second group 
is related to the existence of an exotic fluid with negative pressure such as
Chaplygin gas, $k$-essence, quintessence and others  \cite{alternatives,alternatives2,alternatives3,alternatives4,alternatives5,alternatives6,alternatives7,
alternatives8}. In all these 
alternatives the dark  energy equation of state (EoS) changes dynamically with time differently from the $\Lambda$CDM, the
 EoS of which is constant and equals $-1$. In the context of quintessence, a canonical scalar 
field $\phi$ minimally coupled to gravity is used as a dark energy model. The field $\phi$ varying slowly 
along a potential $V(\phi)$  can lead to  observational results very similar to the cosmological constant 
(for a review see for  example  Refs. ~\refcite{Copeland2006,Tsujikawa2013}). 

One of the most interesting models for inflation and dark energy 
is  an ultra-light  pseudo Nambu--Goldstone boson (pNGB) which is still relaxing to its vacuum state. 
From the viewpoint of quantum field theory, pNGB models are the simplest way to have a naturally 
ultra-low mass,  spin-0 particles and hence, perhaps the most natural candidate for a minimally 
coupled scalar field presently existing. The pNGB effective potential is given by
\begin{equation}
\label{pot1}
V(\phi)=\mu^4\,\left[1 \pm \cos\left(\frac{\phi}{f}\right)\right],
\end{equation}
where $f\sim m_p\simeq 10^{19}\, \textrm{GeV}$ is the Planck scale representing the   
spontaneous breaking scale, and $\mu$ is related to the explicit global symmetry breaking scale.
Note that $f\rightarrow \infty$ corresponds to the exact shift symmetric constant potential. 
pNGBs were first proposed in the context of  natural inflation \cite{nat}, and then they were 
subsequently extended for the case of dark energy \cite{frieman95}. Also, a semiclassical calculation 
of particle production by a Nambu--Goldstone boson is performed in  Ref.~\refcite{Dolgov}. 

On the other hand, analytical integration of the motion equations, in  context of both inflation and quintessence, 
 admits as solutions  potentials  related to hyperbolic functions  \cite{Chaadaev2013, Kim2013, Lobo2013, Espichan2008, Gonzalez2000}. 
In fact, some of these potentials can be considered as hyperbolic extensions of (\ref{pot1}). Additionally,  elliptic functions 
have been considered to study some inflationary aspects  \cite{higaki} and it was recently shown, in  context of  Hamilton-Jacobi method, 
that solutions related to the  inflationary potential can be  expressed in terms of elliptic functions \cite{v15,vg}.  Furthermore, in 
recent times,  Friedman--Robertson--Walker universe (FRW) containing a mixture of perfect fluids and a quintessence escalar field,  admits as solutions
for the potential, some   especial cases
of elliptic functions \cite{Gonzalez2000}. Within this context and keeping in mind that trigonometric and hyperbolic functions are particular cases of 
Jacobi's elliptic functions,  
we propose, in some sense, a unified treatment of all of these potentials. For this reason we generalize Eq. (\ref{pot1}) 
in an elliptic version by means of Jacobi functions\footnote{Jacobi's elliptical functions are not strange in physics and engineering. 
They appear in a variety of problems. For example, they appear when the equation of motion of the bead on a rotating hoop is solved \cite{Baker2012},
 or in the study of the nonlinear Schrodinger equation in the context of Bose-Einstein condensates \cite{Bronski2001}, or when population growth dynamics 
 is studied \cite{Kenkre2003},  as well as in other fields  of physics \cite{Jacob,Jacob2,Jacob3,Jacob4}.}, 
 and we use it to implement  an inflationary era and a late acceleration of the universe. 
Such a generalization owns a shift symmetry, like the pNGB potential, important to solve the two reasons that render challenging to find a 
natural candidate for quintessence field in particle physics.  
The first   reason
is related to the   violation of the flatness condition (see for example Ref.~\refcite{Kolda1998}), while the second  is related to the fact that an ultra-light
field would carry a fifth force, typically of gravitational strength (see for example Refs.~\refcite{Carroll1998},\refcite{Chiba1999}). In this  work we shall limit ourselves
to the case of the FRW homogeneous and isotropic metric  with null curvature.

This paper is organized as follows: in Section \ref{sec:JpNGB} we present the generalized potential which can be 
expressed in terms of elliptic functions. In Section \ref{sec:infJpNGB} an inflationary model based on the generalized 
potential is presented. We provide the general solutions and the expressions for the relevant cosmological 
parameters and then we study the viability of the model in the context of recent data from the {\it Planck collaboration} 2015 \cite{planck15}. 
In Section \ref{sec:DEJpNGB} we study the generalized potential 
as a dark energy model, i.e., as a quintessence model. We perform the study 
of the evolution of its EoS and of the phase space $w'- w$ . Finally,  we perform observational tests 
to investigate the viability of quintessence models 
based on the Jacobian generalization.  In section \ref{concl} we 
present our conclusions and final remarks.

\section{Jacobian pseudo Nambu--Goldstone boson}
\label{sec:JpNGB}
%%%%%%%%%%%%%%%%%%%%%%%%%%%%%%%%%%%%%%%%%%%%%%%%%%%%%%%%%%%%%%%%%%%%%%%%%%%%%%%%%%%%%%%%%%%%%%%%%
Recalling that trigonometric functions are special cases of elliptic functions,
we proceed to generalize the effective potential (\ref{pot1})
in a simple way:

\begin{equation}
\label{pngb1}
V(\phi, \,k)=\mu^4\,\left[1+\textrm{cn} \left(\frac{\phi}{f}\right)\right],
\end{equation}
where $\textrm{cn}(x)\equiv\textrm{cn}(x|\, k)$ is the Jacobi's elliptic cosine function,
and $k$ is the {\it modulus} of the elliptic function. Note that the original pNGB potential is 
recovered by making $k=0$.  We will refer  to potential (\ref{pngb1}) as 
Jacobian pseudo Nambu--Goldstone boson (JpNGB) effective potential. In FIG. \ref{f1} the 
effective potential (\ref{pngb1}) is plotted for different values of the modulus $k$
 and in the Appendix some properties of the Jacobi's elliptic function are presented.
Without loss of generality, we will consider that $\mu$ and $f$ possess the same physical 
meaning as in the  pNGB scheme.

\begin{figure}[h]
  \begin{center}
    \includegraphics[width=80mm]{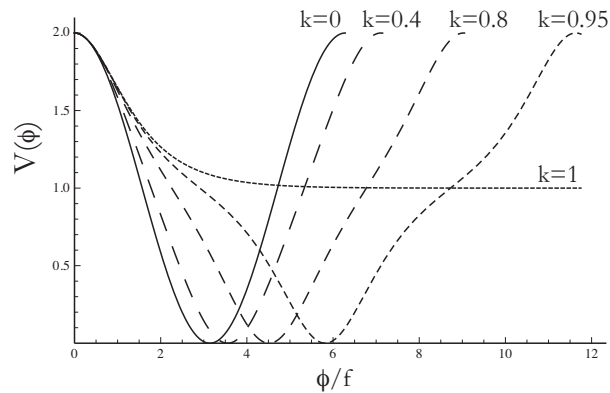}
  \end{center}
  \caption{Plot of the Jacobian pseudo Nambu--Goldstone boson effective potential $V$ as a 
  function of the dimensionless scalar field $\varphi=\phi/f$ for different values 
  of the modulus $k$. Clearly,  the hyperbolic limit $k=1$ does not allow  inflation to end 
  because the inflaton   {\it rolls} eternally.}
  \label{f1}
\end{figure}
Alternatively, we mention that Higaki \& Takahashi \cite{higaki} have recently studied the case with a minus sign in the potential (\ref{pngb1}) calling  this model the 
{\it Cnoidal inflation}. Specifically, they have shown that this model predicts values of the spectral 
index and the tensor--to--scalar ratio that interpolates from natural inflation to exponential  inflation 
such as $R^2$--or Higgs inflation and brane inflation.  In the next section we will use potential (\ref{pngb1}) to implement 
an inflationary epoch in the early universe and we will study its concordance with observational data of the Planck satellite.

\section{JpNGB as a inflaton field}
\label{sec:infJpNGB}
%%%%%%%%%%%%%%%%%%%%%%%%%%%%%%%%%%%%%%%%%%%%%%%%%%%%%%%%%%%%%%%%%%%%%%%%%%%%%%%%%%%%%%%%%%%%%5
In the inflationary regime, the dynamical evolution of the JpNGB field is 
described by the Klein--Gordon equation
\begin{equation}
\label{pngb2}
\ddot{\phi}+(3H+\Gamma ) \dot{\phi}+V_{,\phi}=0,
\end{equation}
where $\Gamma$ is the decay width of the inflation, the
dot indicates the derivative with respect to time, 
$H=\dot{a}/a$ and $V_{,\phi}$ represents the derivative
of the potential with respect to $\phi$.
Note that in the temperature range 
$\mu \lesssim T \lesssim f$, the JpNGB potential
is dynamically irrelevant because the term $V_{,\phi}$
is negligible when compared to the Hubble damping term.
Therefore, in this temperature range, aside from the 
smoothing of spatial gradients in $\phi$, the field does not evolve.
For temperatures $T\lesssim f$, the global symmetry is 
spontaneously broken, and the field $\phi$ describes the 
phase degree of freedom around the bottom of the potential. 
Since $\phi$ thermally decouples at a temperature 
$T \sim f^2/m_p \sim f$
we assume it is initially laid down at random between 0 and $P f$ in 
different causally connected regions, where $P$ is the period of the 
elliptic function.
\begin{figure}[t]
  \begin{center}
    \includegraphics[width=80mm]{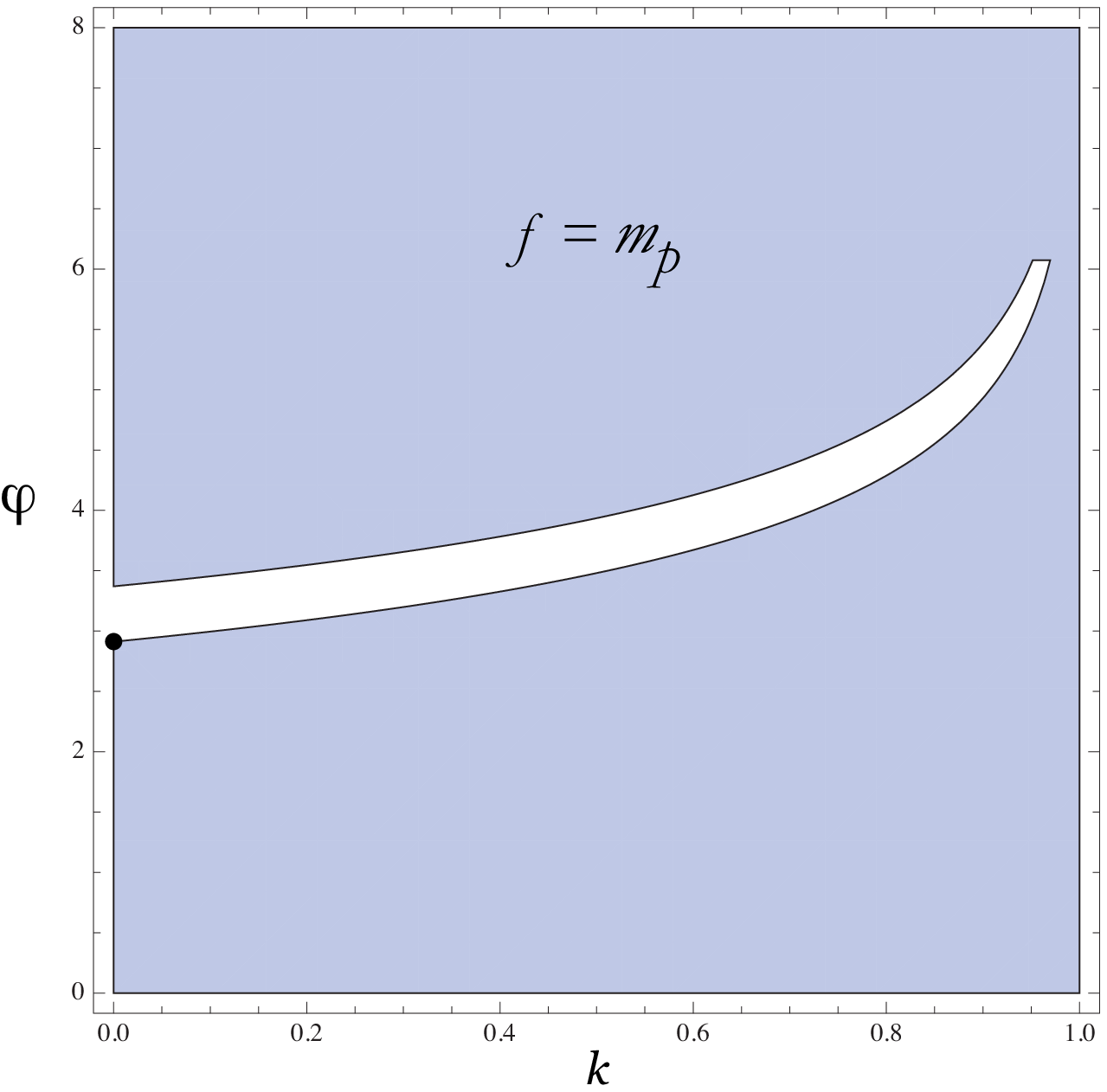}
    \includegraphics[width=80mm]{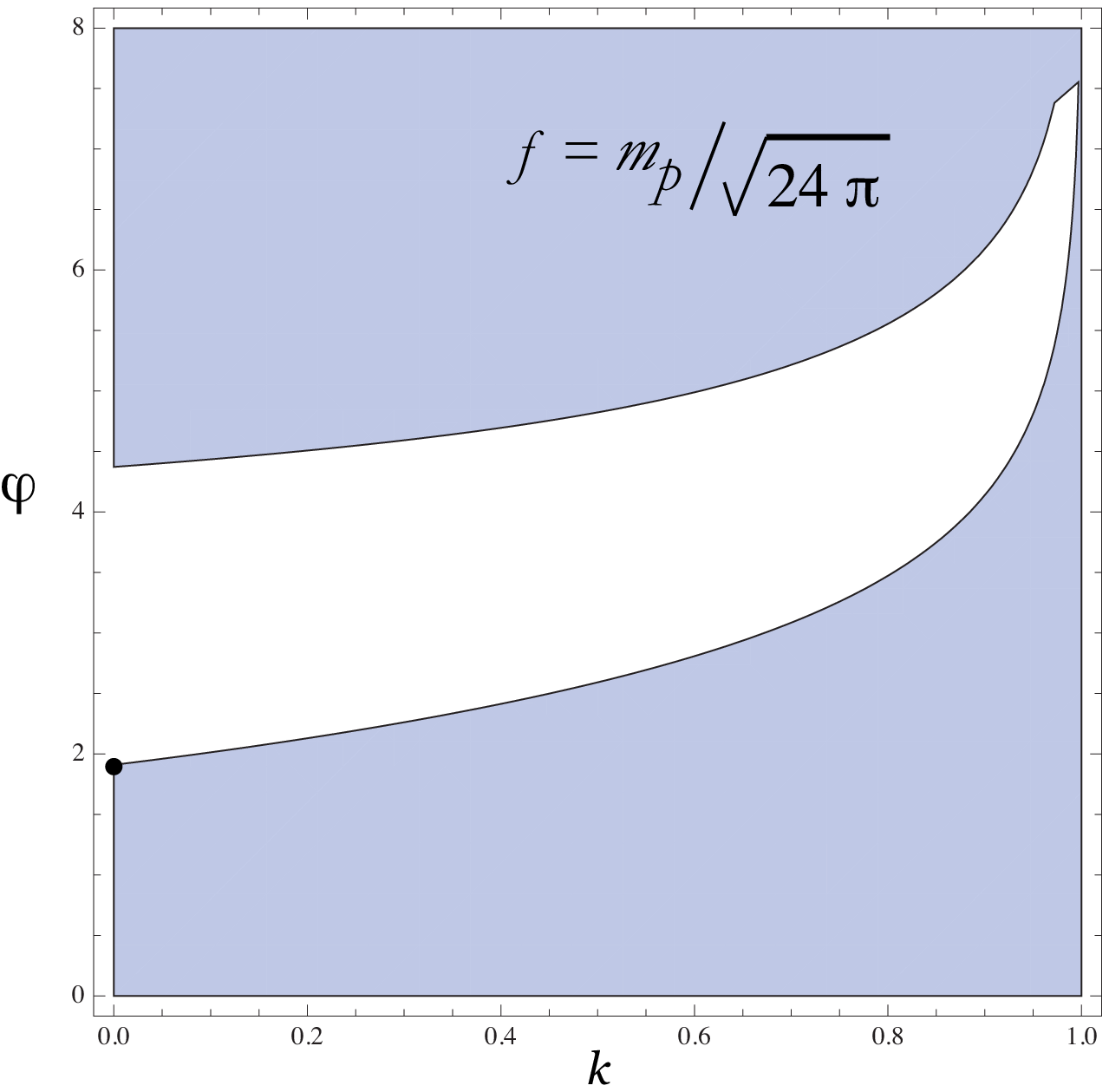}
  \end{center}
  \caption{Plots for the condition (\ref{sr4}) in the $\varphi$--$k$ plane for two
  values of $f$. The
  validity of inequality (\ref{sr4}) is represented by the grey region of each graph and the boundary
  corresponds to the value of dimensionless scalar field $\varphi$ at the end of inflation, $\varphi_e$. 
  Also, the value reported by Freese et al. \cite{nat} for $k=0$, $\varphi_e^{0}$,  is represented by a dot.
  Top panel: $f=m_p$ and $\varphi_e^0=2.98$; Bottom panel: $f=m_p/\sqrt{24 \pi}$ and $\varphi_e^0=1.9$.}
  \label{figcond}
\end{figure}
Finally, at temperatures $T\lesssim \mu$, in regions of the universe with 
$\phi$ initially near the top of the potential, the field starts 
to roll slowly down the hill toward the minimum. 
In those regions, the energy density of the universe 
is quickly dominated by the vacuum contribution 
$[V(\phi) \simeq 2\mu^4 \gtrsim p_{rad}\sim T]$, 
and the universe expands exponentially. The slow-roll   (SR) regime  occurs when the motion 
of the field is overdamped, so 
$\ddot{\phi} \ll 3 H \dot{\phi}$ and $\Gamma \ll H$\footnote{For warm inflation, the inflaton motion is also overdamped 
but under condition 
$\Gamma \gtrsim H$.},
and therefore two conditions are met:
\begin{equation}
\label{sr1}
|V_{,\phi \phi}(\phi| )| \lesssim 9 H^2,
\end{equation}
and 
\begin{equation}
\label{sr2}
\left|\frac{V_{,\phi}(\phi)}{V(\phi)}\right|\lesssim \frac{\sqrt{48 \pi}}{m_p},
\end{equation}
which, using Eq. (\ref{pngb1}) and defining the dimensionless scalar field
$\varphi=\phi/f$, leads to
\begin{equation}
\label{sr3}
\sqrt{\frac{2\, \textrm{cn}\, \varphi\, |\textrm{dn}^2\, \varphi-k^2 \textrm{sn}^2\, \varphi|}{1+\textrm{cn} \,\varphi}}\lesssim \frac{\sqrt{48 \pi} f}{m_p},
\end{equation}
and
\begin{equation}
\label{sr4}
\frac{|\textrm{dn}\,\varphi\, \textrm{sn}\,\varphi|}{1+\textrm{cn}\,\varphi}
\lesssim \frac{\sqrt{48 \pi} f}{m_p},
\end{equation}
respectively (see definitions and relations of Jacobi's elliptic functions in the Appendix). Obviously, the SR regime ends when one of inequalities
(\ref{sr3}) or (\ref{sr4}) is violated. This fact occurs when $\phi$
reaches a value $\phi_{e}$ which is evaluated for $f=m_p$
and $f=m_p/\sqrt{24\pi}$ in FIG. \ref{figcond}. 
Note the consistency between our results and those found previously by 
Freese, Frieman \& Olinto \cite{nat} using k=0, i. e.
$\varphi_e^0=2.98$ for $f=m_p$ and $\varphi_e^0=1.9$ for $f=m_p/\sqrt{24 \pi}$.
\begin{figure}[h]
  \begin{center}
    \includegraphics[width=85mm]{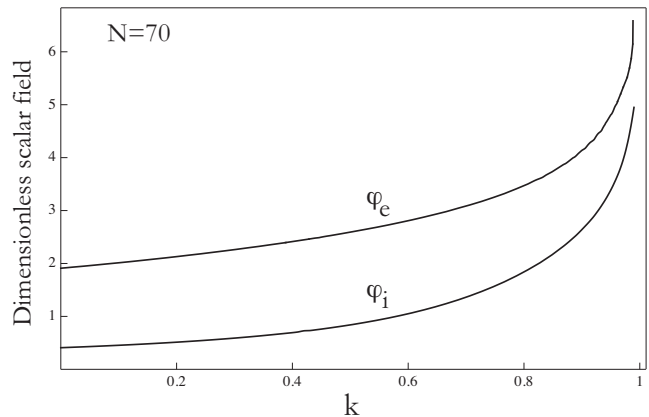}
  \end{center}
  \caption{Numerical solution for the initial value $\varphi_i$ and final value
  $\varphi_e$ of the inflaton as a function of the modulus $k$ using $N=70$ and
  $f=m_p/\sqrt{24 \pi}$.}
  \label{fign}
\end{figure}

 Now let us calculate the values of the inflaton related to the number of e-folds required 
to generate the seeds for the creation of the large scale structures. As is well known, the number of e-folds 
of physical expansion that occur in the inflationary stage is given by
\begin{eqnarray}\nonumber
N(\phi)&=&-\frac{8 \pi}{m_p^2}\int_{\phi}^{\phi_e}\frac{V}{V_{,\phi}}d\phi\\ 
&=& \frac{8 \pi f^2}{m_p^2}\ln\left(\frac{\textrm{ds}\, \varphi}{\textrm{ds}\, \varphi_e}
\sqrt{\frac{1-\textrm{cn}\, \varphi_e}{1+\textrm{cn}\, \varphi_e}
\frac{1+\textrm{cn}\, \varphi}{1-\textrm{cn}\, \varphi}}\,e^{\Phi -\Phi_{e}}\right),\label{N}
\end{eqnarray}
where 
\begin{equation}
\label{fgg}
\Phi \equiv \sqrt{\frac{k}{k'}}\,\arctan\left(\sqrt{\frac{k}{k'}}\,\textrm{cn} \,\varphi\right),
\end{equation}
and $k'=1-k$ is the complementary modulus. Therefore, knowing the value of
$\phi_e$ from Eq. (\ref{sr4}) the initial value of the inflaton $\phi_i$ can be calculated,
imposing a value on $N$, for example, $N(\phi_i)=70$. In FIG. \ref{fign} we show
the initial $\varphi_i$ and final $\varphi_e$ values of inflaton calculated using Eq.  (\ref{N})
and $f=m_p/\sqrt{24\pi}$ as a function of the modulus $k$.  Our results indicate that for 
any value of $k$, the inflaton with a JpNGB effective potential is compatible with 70 e-folds. We have 
checked for values of $N(\phi_i)$ between 60 and 70 and have found that the compatibility is maintained. 
\begin{figure}[h]
  \begin{center}
    \includegraphics[width=85mm]{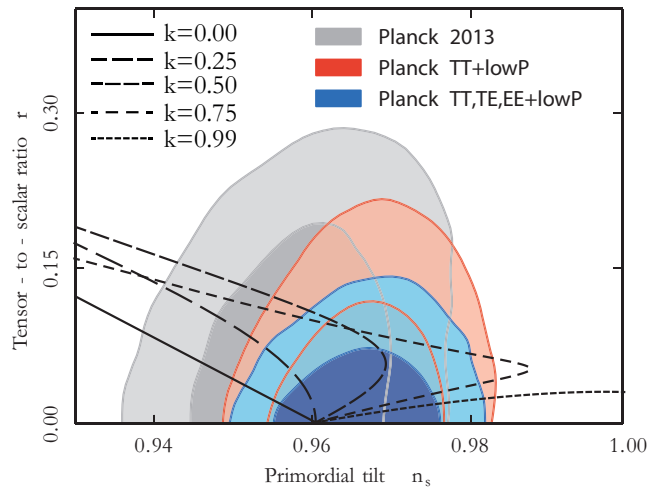}
  \end{center}
  \caption{Plot of the tensor--to--scalar parameter $r$ as a function of the scalar spectral index $n_s$ for the
Jacobian pseudo Nambu--Goldstone Bosons potential for  $f=m_p/\sqrt{24 \pi}$. The marginalized joint $68\%$ 
and $95\%$ confidence level regions using Planck TT + low P, Planck TT, TE, EE + low P and Planck 2013 data
release with $\widehat{ \rm{k}}=
 0.05h$ Mpc$^{-1}$ are shown  \cite{planck15}.}
  \label{figpl}
\end{figure}

In the inflationary scenario, the quantum fluctuations are  relevant because they 
generate two important types of perturbations: density perturbations (arising from
quantum fluctuations in the scalar field, together with the corresponding scalar metric perturbation),
and relic gravitational waves (which are tensor metric fluctuations).  The former is sensitive to 
gravitational instability and leads to structures formation, while the latter predicts a stochastic 
background of relic gravitational waves. Scalar perturbations produce a power spectrum characterized
by the scalar spectral index $n_s$ while tensorial perturbations produce a power spectrum characterized
by the gravitational wave spectral index $n_T$. In inflationary models it is common to define the tensor--to--scalar 
amplitude $r =\frac{P_T}{P_R}$, where $P_T$ and $P_R$ are amplitudes of the tensor and scalar power spectrum 
respectively.

The spectral index $n_s$ and the tensor--to--scalar amplitude ratio $r$ are written  in terms 
of the slow--roll parameters,
\begin{eqnarray}
\label{eps1}
\epsilon_V&=&\frac{m_p^2}{16\pi} \left(\frac{V_{,\phi}}{V}\right)^2=\frac{m_p^2}{16\pi f^2}\,
\frac{\textrm{dn}^2\, \varphi \, \textrm{sn}^2\, \varphi}{(1+\textrm{cn}\, \varphi)^2} \\ \label{eta1}
\eta_V&=&\frac{m_p^2}{8\pi} \left(\frac{V_{,\phi \phi}}{V}\right)=\frac{m_p^2}{16\pi f^2}\,
\frac{\textrm{cn}\, \varphi\,(2\, k\, \textrm{sn}^2\, \varphi-1)}{1+\textrm{cn}\, \varphi},
\end{eqnarray}
the spectral index and the tensor--to--scalar ratio are given by
\begin{eqnarray}\nonumber
n_s &\simeq&1+2\,\eta_V-6\,\epsilon_V\\ 
&=&1-\frac{3 m_p^2}{8\pi f^2}\,\frac{\textrm{dn}^2\, \varphi \, \textrm{sn}^2\, \varphi}{(1+\textrm{cn}\, \varphi)^2}+\frac{m_p^2}{4\pi f^2}\,\frac{\textrm{cn}\, \varphi\,(2\, k\, \textrm{sn}^2\, \varphi-1)}{1+\textrm{cn}\, \varphi},\\ \label{ns1}
r &\simeq& 16\,\epsilon_V=\frac{m_p^2}{\pi f^2}\,\frac{\textrm{dn}^2\, \varphi \, \textrm{sn}^2\, \varphi}{(1+\textrm{cn}\, \varphi)^2}.\label{r1}
\end{eqnarray}
Therefore,  using $n_s$ and $r$, it is possible 
to obtain a direct comparison of our results with 
the {\it Planck collaboration} data sets, 
which is shown in Figure \ref{figpl}. 
Based on  such comparisons, we may say 
that a description of the inflationary universe
model in terms of a JpNGB effective potential 
could quite well accommodate the data recently released 
by the Planck mission  \cite{planck15}. 
However, potentials with $k\sim 1$  are not viable because 
the inflaton rolls eternally.

\section{JpNGB as dark energy}
\label{sec:DEJpNGB}
%%%%%%%%%%%%%%%%%%%%%%%%%%%%%%%%%%%%%%%%%%%%%%%%%%%%%%%%%%%%%%%%%%%%%%%%%%%%%%%%%%
It is well known that a canonical scalar field $\phi$, called quintessence, varying slowly along  
one potential $V(\phi)$ can lead to observational results very similar to the   cosmological 
constant and would explain the late 
acceleration  of the universe  (for a review see for example Ref.~\refcite{Tsujikawa2013}). 
However, two reasons render it challenging to find a natural
candidate for $\phi$ in particle physics. The first  is related to the violation of the flatness
condition \cite{Kolda1998}, while the second  is related to the fact that an ultra-light
field would carry a fifth force, typically of gravitational strength \cite{Carroll1998, Chiba1999}.
These problems can be solved by means of a potential which has a shift symmetry, as the pNGB potential
\cite{frieman95}.

The fact that potential (\ref{pngb1}), which is a generalization of pNGB, is a periodic 
function and therefore has a shift symmetry motivates us regarding the possibility of using it in order
to construct a  viable quintessence model. Thus we exploit  its dynamic behavior 
when it is used   to explain accelerated expansion in late times.  

We start by considering the general motion  equations  in the case where the universe  is dominated by a 
pressureless matter and a quintessence field. In  the context of  a  homogeneous and isotropic 
Friedmann-Robertson-Walker universe, with null curvature and in the case where non-relativistic matter
is not interacting with the quintessence field we have
\begin{eqnarray}\label{motion}
\ddot{\phi}+3H \dot{\phi}+V_{,\phi}=0  \\
\dot{\rho}_m+3H\rho_m=0   \\
\dot{\rho}_\phi+3H(\rho_\phi+p_\phi)=0   \\ 
\label{motion2}
3H^2m_P^2=\rho_m+\rho_\phi \,.
\end{eqnarray}
where $\rho_\phi = \frac{1}{2}\dot{\phi}+V(\phi)$ and $p_\phi = \frac{1}{2}\dot{\phi}-V(\phi)$  are
the density and pressure of the quintessence field respectively, and $w=p_\phi/\rho_\phi$
is the EoS of dark energy and subindex $m$ indicates matter. From (\ref{motion})-(\ref{motion2}) 
equations  for $w$ and $\Omega_\phi=\rho_{\phi}/\rho_{crit}$, where  $\rho_{crit}=3H_0^2m^2_p$ 
can be written (see for example Refs. ~\refcite{Dutta2008,Scherrer2008})
\begin{eqnarray} \label{equations}
 w'=(1-w)\left[-3(1+w)+\lambda\sqrt{3(1+w)\Omega_\phi}\right] \\
 \Omega'_\phi=-3w\Omega_\phi(1-\Omega_\phi) \,.
\end{eqnarray}
Additionally, one equation for the auxiliary variable $\lambda=-m_p V_{,\phi}/V$
is found
\begin{eqnarray}\label{lambda}
 \lambda'=-\sqrt{3(1+w)\Omega_{\phi}} \lambda^2 (\widetilde{\Gamma} -1) \,,
\end{eqnarray}
where $\widetilde{\Gamma}=V_{,\phi\phi}\,V/V_{,\phi}^{2}$ and $'$ indicates derivative with 
respect to $\ln a$. Thus, given a potential $V$, $\widetilde{\Gamma}$ is known and the system 
(\ref{equations})-(\ref{lambda}) is completely determined. For example, for the exponential 
potential $\widetilde{\Gamma}=1$ and $\lambda$ is a constant and for pNGB potential 
$\tilde{\Gamma}=(\lambda^2-1)/2\lambda^2$. For our case $\tilde{\Gamma}$ has a more
general dependence of $\lambda$ and we will find it numerically.

Recently,  it  was shown that  quintessence models can be approximately  separated  into two classes, 
depending on the acceleration or deceleration of  the field as it evolves down its potential
towards to a zero minimum  \cite{Caldwell2005}. Such groups are called thawing and 
freezing quintessence and they behave quite differently. Models where the field is nearly frozen
away from the minimum  of the potential due to large Hubble dumping during the early cosmological 
era and starts to evolve towards the minimum recently are called thawing. Models  where the 
field initially rolls towards its potential minimum and has only recently gradually slow down are called 
freezing \cite{Caldwell2005}. Examples of the first group are the exponential and  pNBG potentials
\cite{frieman95,thawing, Scherrer2008,Clemson2009,Sen2010,Bellido2011}  and examples of the second group are the inverse power law potentials 
\cite{Dutta2011,freezing,freezing2,freezing3}. In the next subsection we will show that our generalization belongs to  the
thawing group.
\begin{figure}[t]
  \begin{center}
    \includegraphics[width=80mm]{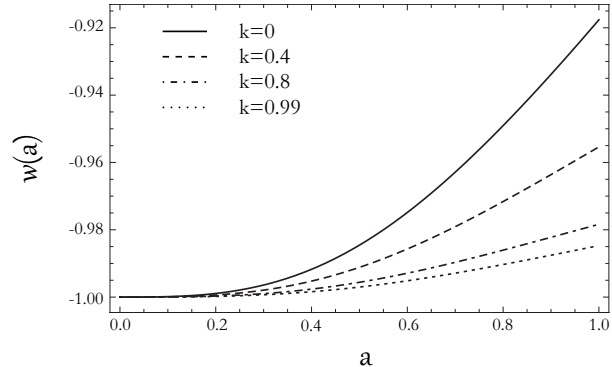}
  \end{center}
  \caption{Evolution of EoS $w$ as a function of scale factor $a$ for different values 
  of the parameter $k$ but same initial  conditions.  The case $k=0$ represents  pNGB potential.
  }
  \label{f2}
\end{figure}
%%%%%%%%%%%%%%%%%%%%%%%%%%%%%%%%%%%%%%%%%%%%%%%%%%%%%%%%%%%%%%%%%%%%%%%%%%%%%
\subsection{JpNGB as a thawing quintessence}
%%%%%%%%%%%%%%%%%%%%%%%%%%%%%%%%%%%%%%%%%%%%%%%%%%%%%%%%%%%%%%%%%%%%%%%%%
As noted above, in thawing quintessence models,  the field is initially frozen in 
early times and then as $H$ increases the field rolls down the potential.
This means that at early times the EoS is $w\approx -1$ but grows less
negative with time where $w'>0$.   Since we are generalizing the quintessence 
potential by means of (\ref{pngb1}), we are interested in knowing if such 
thawing behavior appears in this particular case.

\begin{figure}[h]
  \begin{center}
    \includegraphics[width=80mm]{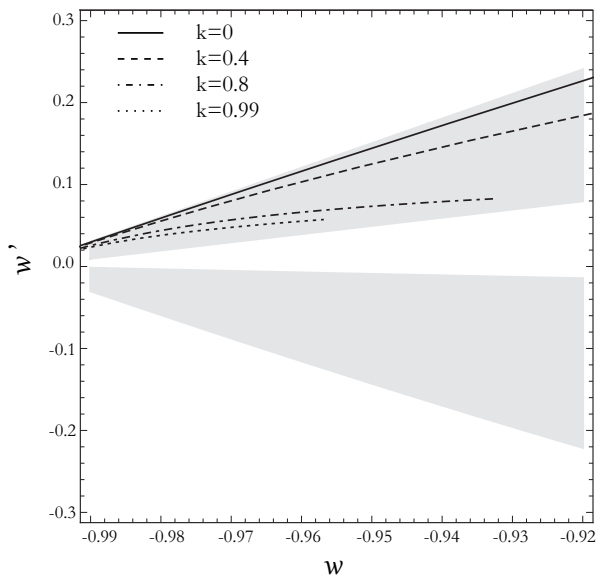}
  \end{center}
  \caption{Plane representing the dynamical space phase $w'-w$  and curves for  JpNGB  effective potential 
   for different values of the modulus $k$. The upper shadowed area is the thawing region $1+w \lesssim w'<3(1+w)$  
and the lower  shadowed area represents  freezing region $0.2w(1+w) \lesssim w'<3w(1+w)$.}
  \label{f3}
\end{figure}

In order to search the  thawing characteristic we solve numerically the system 
(\ref{equations})-(\ref{lambda}) with same initial conditions and for different values 
of $k$-parameter. Results are shown in Figures (\ref{f2}) and (\ref{f3}). In FIG. (\ref{f2}) 
the evolution of the EoS as a function of scale factor $a$ is presented. For all cases we  have used 
the same conditions: $f/m_p= 0.9$ and  $\phi_i/f=1.5$. As we can see, the same initial conditions lead 
us to different evolutions  of EoS for each  $k$. This opens the possibility of discriminating 
between different members of the family potentials (\ref{pngb1}) using observational data. 
In all cases $w$ is close to $-1$ for early times but begins to evolve for late times. Of course $k=0$ 
represents  pNGB potential. We can  note also that in the past, models with potential (\ref{pngb1}) 
are indistinguishable from  the pNGB model and differences between them appear only recently.
 Additionally, it can be seen that the evolution of $w$ is slower for higher values of $k$.

In FIG. (\ref{f3}) we plot the behavior of the model on the dynamical phase space $w'-w$ 
for different values of $k$ and for the same initial  conditions. 
The upper shadowed  area  represents the thawing region $1+w \lesssim w'<3(1+w)$  
and the lower  shadowed area represents the freezing region $0.2w(1+w) \lesssim w'<3w(1+w)$ 
\cite{Caldwell2005}. The values used in FIG.(\ref{f3}) are $f/m_p =1.1$ and $\phi_i/f =1.5$. Note that in all 
cases, curves are inside the thawing region and  we can thus infer that JpNGB has thawing quintessence features.
We have checked for a wide range of initial conditions ($0.6<\phi_i/f<2.5$) and $k$-parameters ($0<k<1$) 
and such thawing features are maintained. Thus we would expect that thawing features are always maintained.

\begin{figure}[h]
  \begin{center}
    \includegraphics[width=80mm]{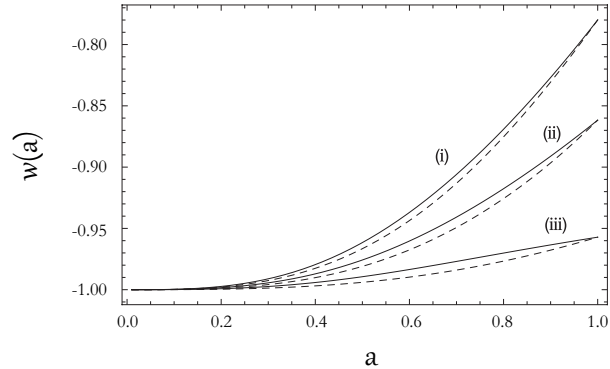}
  \end{center}
  \caption{Evolution of the EoS as a function of $a$ using   
exact and  approximate solutions. Dashed curves represent
exact numerical solutions while continuous curves represent approximate solutions.}
  \label{f4}
\end{figure}

Interestingly, for models having freezing and thawing properties an approximation of $w$ can be known analytically.
In Refs. ~\refcite{Dutta2008,Chiba2009b} the equations of motion (\ref{motion})  were solved in the limit  when $w$ is close to $-1$ 
and an approximated EoS as a function of the scale factor was found when $|w+1|\ll1$. The former condition is ensured if 
conditions of slow roll thawing quintessence $\frac{V'^2}{6H^2V}\ll 1$ and $\frac{V''}{3H^2}\ll1$ \cite{Chiba2009b}, in 
the case where both matter and a scalar field contribute to the density are satisfied. Such approximation gives us
(see Refs.~\refcite{Dutta2008,Chiba2009b} for details)

\begin{eqnarray}\label{eos}
w(a)=-1+(1+w_0)a^{3(K-1)}\times \nonumber\\ 
\left(\frac{(K-F(a))(F(a)+1)^K+(K+F(a))(F(a)-1)^K}{(K-\Omega^{-1/2}_{\phi 0})(\Omega^{-1/2}_{\phi 0}+1)^K+(K+\Omega^{-1/2}_{\phi 0})(\Omega^{-1/2}_{\phi 0}-1)^K}\right)^2
\end{eqnarray}
where $w_0$ is the EoS in the current time and $K$ measures the potential curvature and it is 
\begin{eqnarray} \label{ka}
K=\sqrt{1-\frac{4M^2_p}{3}\frac{V_{,\phi \phi}(\phi_i)}{V(\phi_i)}} \,,
\end{eqnarray}
with $\phi_i$ being the initial value for $\phi$ and function $F(a)$ has
the form
\begin{eqnarray}
F(a)=\sqrt{1-(\Omega^{-1}_{\phi 0}-1)a^{-3}} \,.
\end{eqnarray}
Such approximation was found by means of a Taylor expansion around
$\phi_i$ \cite{Chiba2009b} and comparisons with other approximations 
like Chevallier-Polarski-Linde \cite{Chevallier2003} and Crittenden \cite{Crittenden2007} 
were performed in  Ref.~\refcite{Chiba2009b}. 
As shown in Refs.~\refcite{Dutta2008,Chiba2009b}, numerical solutions are 
in excellent  agreement with the approximate $w$ for several quintessence
models including pNGB. We made the same comparison between exact and approximated EoS for  
JpNGB quintessence models. Results are presented in  FIG.\ref{f4} for different values of $k$. We  have used the 
following values (i) $k=0.4$ $f/m_p=2$, $\phi_i/f=3$ ($K=1.6$); (ii) $k=0.8$, $f/m_p=2$ and 
$\phi_i/f=4$ ($K=1.6$) and  (iii) $k=0.99$, $f/m_p=2$ and $\phi_i/f=6.6$ ($K=1.1$). In all 
figures, continuous curves correspond to an approximate solution (Eq,(\ref{eos})) and dashed 
curves represent  exact numerical solutions of the system (\ref{equations})--(\ref{lambda}).

\begin{figure}[t]
  \begin{center}
    \includegraphics[width=80mm]{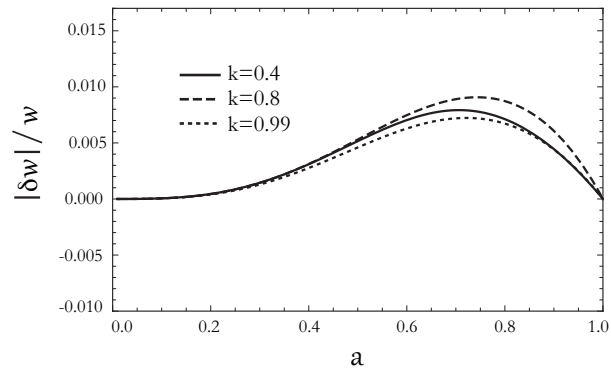}
  \end{center}
  \caption{Curves with relative error as a function of scale parameter $a$ for  curves (i), (ii)
  and (iii)  of FIG.(\ref{f4}). Note that relative error is at most  $\lesssim 0.9\%$.}
  \label{f5}
\end{figure}

In order to evaluate the relative difference between  exact and approximate solutions, 
the relative error ($|\delta w/w|$) was constructed for the curves in FIG.(\ref{f4}). Results
are presented in FIG. (\ref{f5}). Note that in all cases the error is $\lesssim 0.9 \%$, 
which is in agreement with relative errors for other quintessence models  \cite{Chiba2009b}.
We have examined several cases by varying initial conditions, $k$ parameter, etc. and in all of 
them the highest relative error was $\sim 1\%$.  As a conclusion, Eq. (\ref{eos}) gives a good approximate solution of EoS for our 
generalized model.  Parametrization (\ref{eos}) lead us to  $\Omega_{\phi 0}$, $w_0$ and $K$ as free parameters
for EoS.  In the next section we will use (\ref{eos}) to study observational viability of   JpNGB quintessence models
and  constrain free parameters  by using recent observational data. Additionally we will put constraints
on the $k$-parameter.

%%%%%%%%%%%%%%%%%%%%%%%%%%%%%%%%%%%%%%%%%%%%%%%%%%%%%%%%%%%%%%%%%%
\subsection{Observational viability}
%%%%%%%%%%%%%%%%%%%%%%%%%%%%%%%%%%%%%%%%%%%%%%%%%%%%%%%%%%%%%%%%
Analysis considering Eq. (\ref{eos}) for  thawing models   have already been  carried out in 
Refs. ~\refcite{Dutta2008,Chiba2009c,Chiba2013}.  For example in Ref.~\refcite{Chiba2009c} the likelihood 
analysis with the SNIa and BAO data was performed. In Ref. ~\refcite{Chiba2013} these analyses were updated 
by using the Union 2.1 data set \cite{Suzuki2012}, the CMB shift parameter measured by WMPA7 \cite{Wmap} 
and the BAO data of the BOSS experiment \cite{boss}. 

In this subsection we present an updated observational analysis by using the latest SNIa data (JLA sample \cite{Betoule2014})
and $H(z)$ data \cite{Farooq2013}. Also, a first attempt to apply Eq. ({\ref{eos}}) to study  JpNGB models
is developed. Our  tests are based on $\chi^2$-statistics, which will allow us to explore 
the space of parameters  and constrain the free parameters $\theta$. 

Function $\chi^2$ is given by
\begin{equation}
\chi^{2}(\theta)=\Delta y ^T (\theta)\mathbf{C}^{-1} \Delta y(\theta)\,,
\label{chi2}
\end{equation}
where $\Delta y(\theta) = y_i-y(x_i;\theta)$, $\mathbf{C}$ is the covariance matrix of data points $y_i$ and
$y(x_{i}\vert\theta)$  represents   theoretical predictions 
for a given set of parameters. The best fit is found by minimizing the $\chi^2$-function and   
the minimum of $\chi^2$ gives us an indication of the quality of the fit.

First, we considered  tests involving the distance modulus of type Ia Supernovae, 
which is defined by
\begin{eqnarray}
 \mu(z,\theta)=5\,log_{10}(d_L(z,\theta))+42.38-5\,log_{10} h \,, 
\end{eqnarray}
where the luminosity distance is given by
\begin{eqnarray}
 d_L(z,\theta)=(1+z)H_0\int^{z}_0\frac{dz'}{H(z',\theta)} \,.
\end{eqnarray}
For our observational treatment with SNIa we  used the  JLA sample 
\cite{Betoule2014} which has 740 data points. This updates a previous analysis 
performed in \cite{Chiba2013} based on the Union2.1 data set  \cite{Suzuki2012}.
Observational data points of the luminosity-distance  modulus were calculated
using relation \cite{Betoule2014}
\begin{equation} \label{muobs}
\mu_{obs} = m^*_B-(M_B-\alpha X_1+\beta C)\,,
\end{equation}
where $m^*_B$ corresponds to the observed peak magnitude in the rest frame $B$ band and $\alpha$, $\beta$
and $M_B$ are nuisance parameters, $X_1$ is related to the time stretching of the light curves,
and $C$ corrects the color at maximum brightness. In order to calculate completely $\mu_{obs}$
and its covariance matrix we have followed the steps suggested in Ref.~\refcite{Betoule2014}. 

 After that, we performed an analysis by using  the compilation 
of observational points for $H(z)$ parameter, which were derived  using the 
differential evolution of passively evolving  galaxies  as cosmic chronometers 
\cite{Simon2005,Stern2010}. We used the recently updated $H(z)$ data \cite{Farooq2013} 
and finally we performed an joint analysis  using $\chi^2_{total}=\chi^2_{SNIa}+\chi^2_{H(z)}$.

\begin{figure*}% Inicia o ambiente de figuras
\centering
  \subfloat[]{ % ComeÃ§a a incluir a figura fig1.pdf
   \includegraphics[width=5.6cm,height=4.3cm]{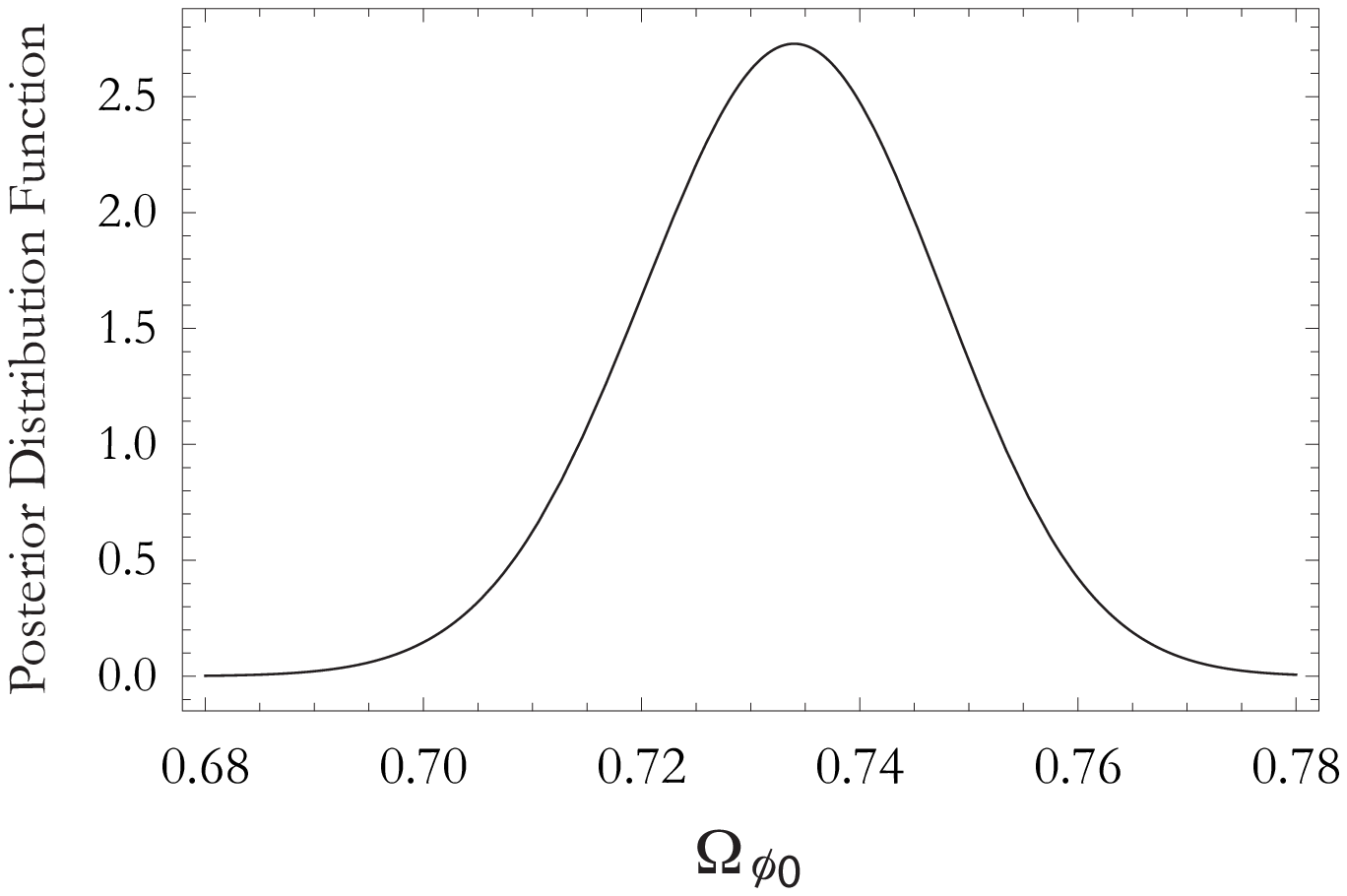}
  } % Termina de incluir a figura fig1.pdf
  \subfloat[]{ % ComeÃ§a a incluir a figura fig2.pdf na mesma linha da figura fig1.pdf
    \includegraphics[width=5.6cm,height=4.3cm]{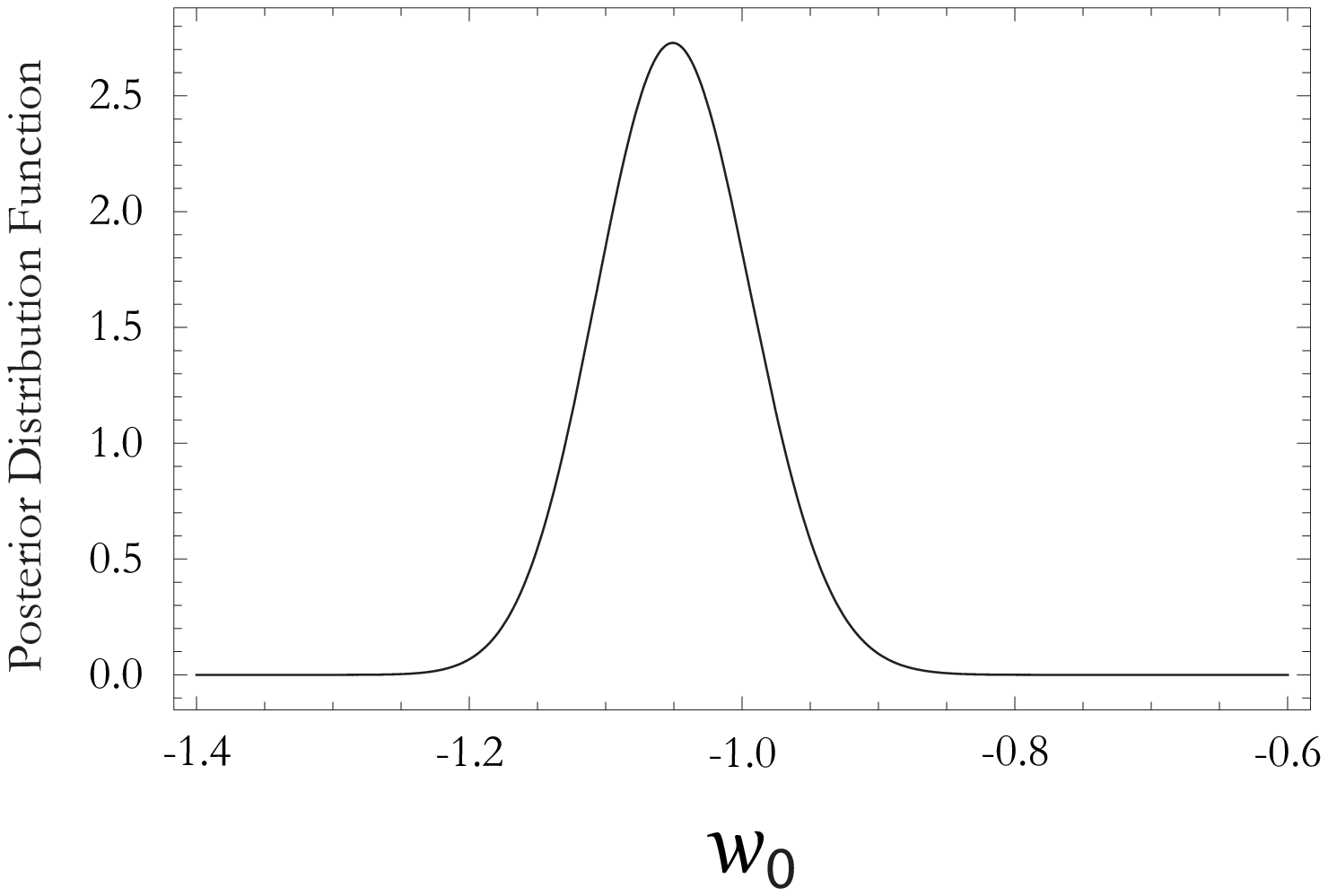} 
  } \\% Termina de incluir a figura fig2.pdf
  
  \subfloat[]{ % ComeÃ§a a incluir a figura fig2.pdf na mesma linha da figura fig1.pdf
    \includegraphics[width=5.6cm,height=4.3cm]{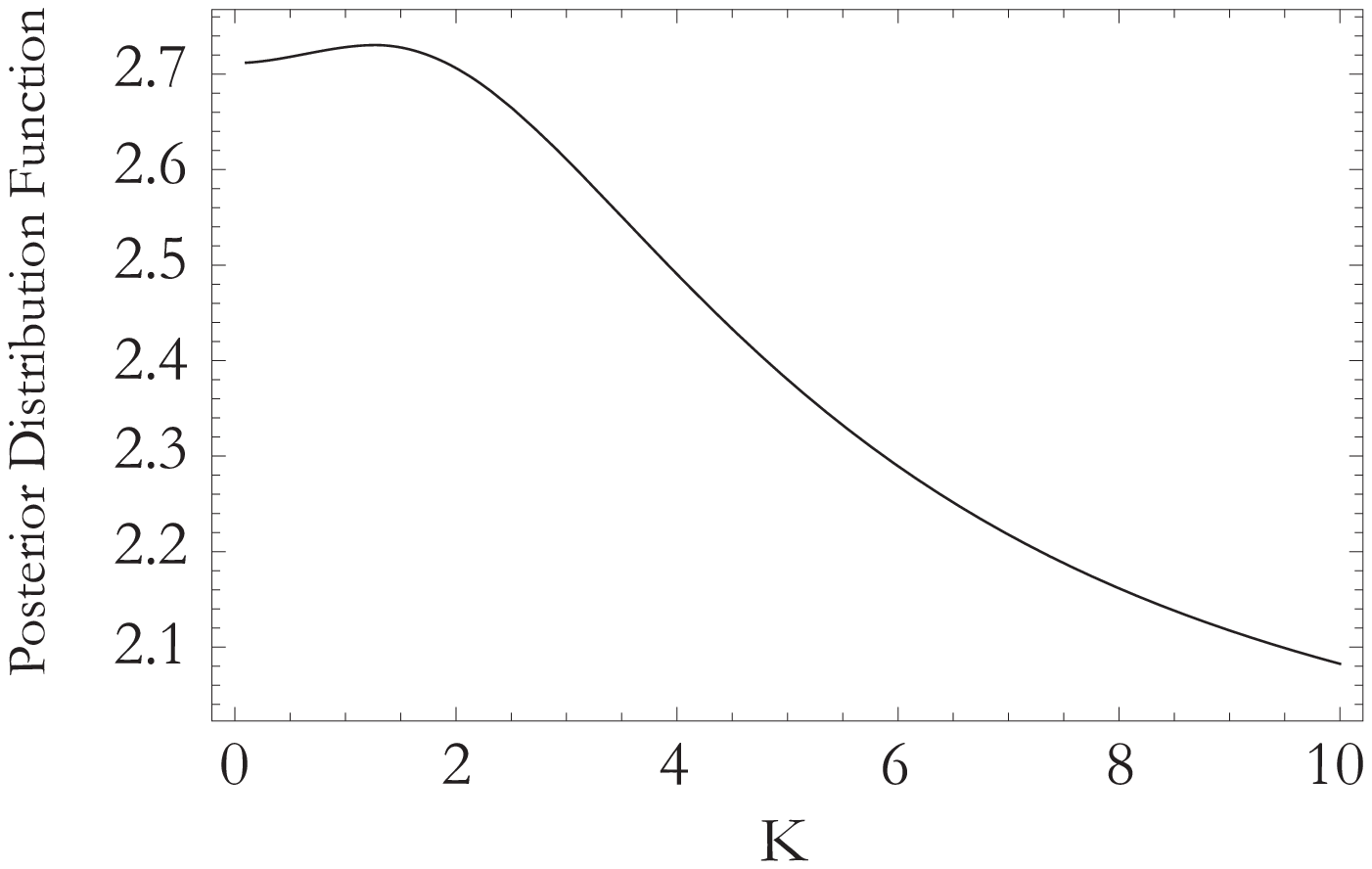}
  } % Termina de incluir a figura fig2.pdf
   % Com esse comando iremos incluir a Ãºltima figura na prÃ³xima linha
  \subfloat[]{ % ComeÃ§a a incluir a figura fig3.pdf na linha abaixo
    \includegraphics[width=5.2cm,height=4.7cm]{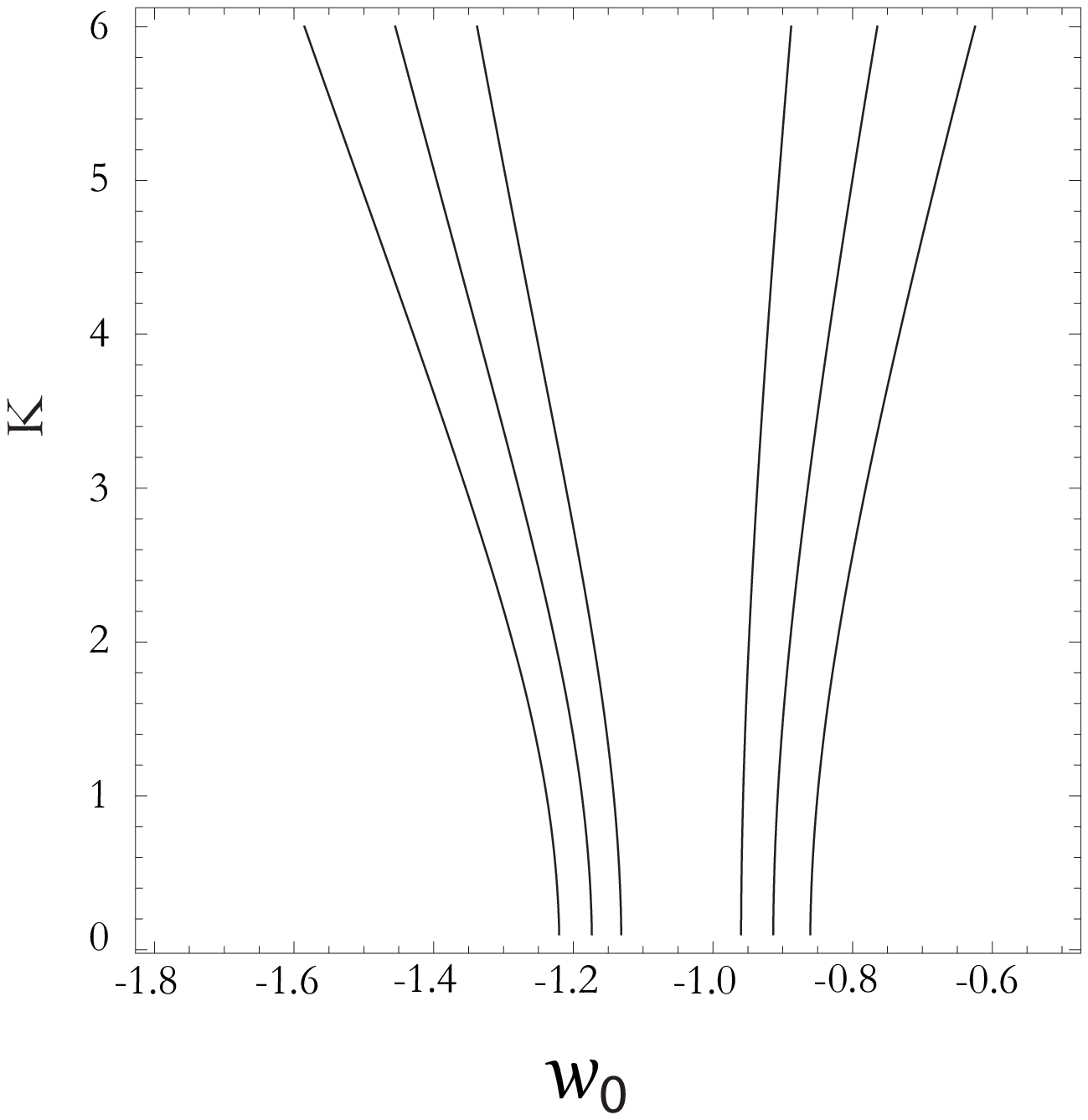}
  } 
 \\
  \caption{(a) PDF function for parameter $\Omega_{\phi 0}$ 
after marginalization on $w_0$ and $K$, (b) PDF function for parameter $w_{0}$  
after marginalization on $\Omega_{\phi 0}$ and $K$, (c)  PDF function of
parameter $K$ after marginalization on $\Omega_{\phi 0}$ and $w_0$ and
(d) $w_0$ - $K$ plane with contour lines at 1$\sigma$, 2$\sigma$ and 3$\sigma$ 
CL.}
\label{f6}
  \end{figure*} % Fecha o ambiente de figuras

Considering Eq (\ref{eos}) in an observational context of $\mu(z,\theta)$ 
and $H(z)$ leads us having 4 free parameters ($h$, $\Omega_{\phi 0}$, $w_0$ and $K$).
However,  we have marginalized over parameter $h$ with the prior 
$ 50 km \,s^{-1} Mpc^{-1} \leq H_0\leq 80 km \,s^{-1} Mpc^{-1}$
in all of our analyses. Thus, we leave   all other three  
parameters free. As discussed in \cite{Chiba2013} and for reasons related
to  approximation limits of Eq. (\ref{eos}), we used  a 
prior  $0<K<10$. Using SNIa alone,  our results give us the 
constraint of $\Omega_{\phi 0}=0.73^{+0.06}_{-0.07}$, $w_0 = -1.13^{+0.15}_{-0.17}$
and $K=0.99^{+8}_{-0.99}$ ($2\sigma$ CL). Using $H(z)$ alone 
our results give us the  constraints for $\Omega_{\phi 0}=0.72^{+0.05}_{-0.05}$, $w_0 = -0.80^{-0.30}_{+0.35}$
and $K=0.1^{+8.5}_{-0.1}$ ($2\sigma$ CL), while the joint analysis gives us
$\Omega_{\phi 0}=0.73^{+0.04}_{-0.05}$, $w_0 = -1.05^{+0.13}_{-0.17}$
and $K=0.99^{+8.2}_{-0.99}$ ($2\sigma$ CL) with a $\chi^2_\nu \sim 1.1$. 
Results of the joint analysis are presented in FIG. (\ref{f6}). 
FIG. \ref{f6} (a) shows us the PDF function of the parameter $\Omega_{\phi 0}$ 
after marginalization over $w_0$ and $K$. FIG. \ref{f6} (b) shows us the 
PDF function for parameter $w_{0}$  after marginalization on $\Omega_{\phi 0}$ and $K$.
FIG. \ref{f6} (c) shows us the PDF function for parameter $K$ 
after marginalization on $\Omega_{\phi 0}$ and $w_0$. Finally 
FIG. \ref{f6} (d) visualizes the $w_0$ - $K$ plane with 
contour lines at 1$\sigma$, 2$\sigma$ and 3$\sigma$ confidence levels.
Our results here are in agreement with results in \cite{Chiba2013}.
It can be seen  that there is a very high degeneration for parameter $K$.

So far,  we have considered the parameter $K$. However, in general, this 
parameter is not adequate to discriminate between different thawing models 
through the data. Since we are interested in knowing if JpNGB models are
preferred or ruled out by the observations, it is necessary to consider $K$  as
a function of parameter $k$. Actually, $K$ is dependent on of $\phi_i/f$ also.
However, since $\phi_i/f$ and $k$ are poorly constrained by data considered here,
we will fix the value of $\phi_i/f$ and vary the other three parameters. This will 
allow us to have information about the remaining parameters, particularly of the $k$ parameter.

\begin{figure*}% Inicia o ambiente de figuras
\centering
  \subfloat[]{ % ComeÃ§a a incluir a figura fig1.pdf
   \includegraphics[width=5.6cm,height=4.5cm]{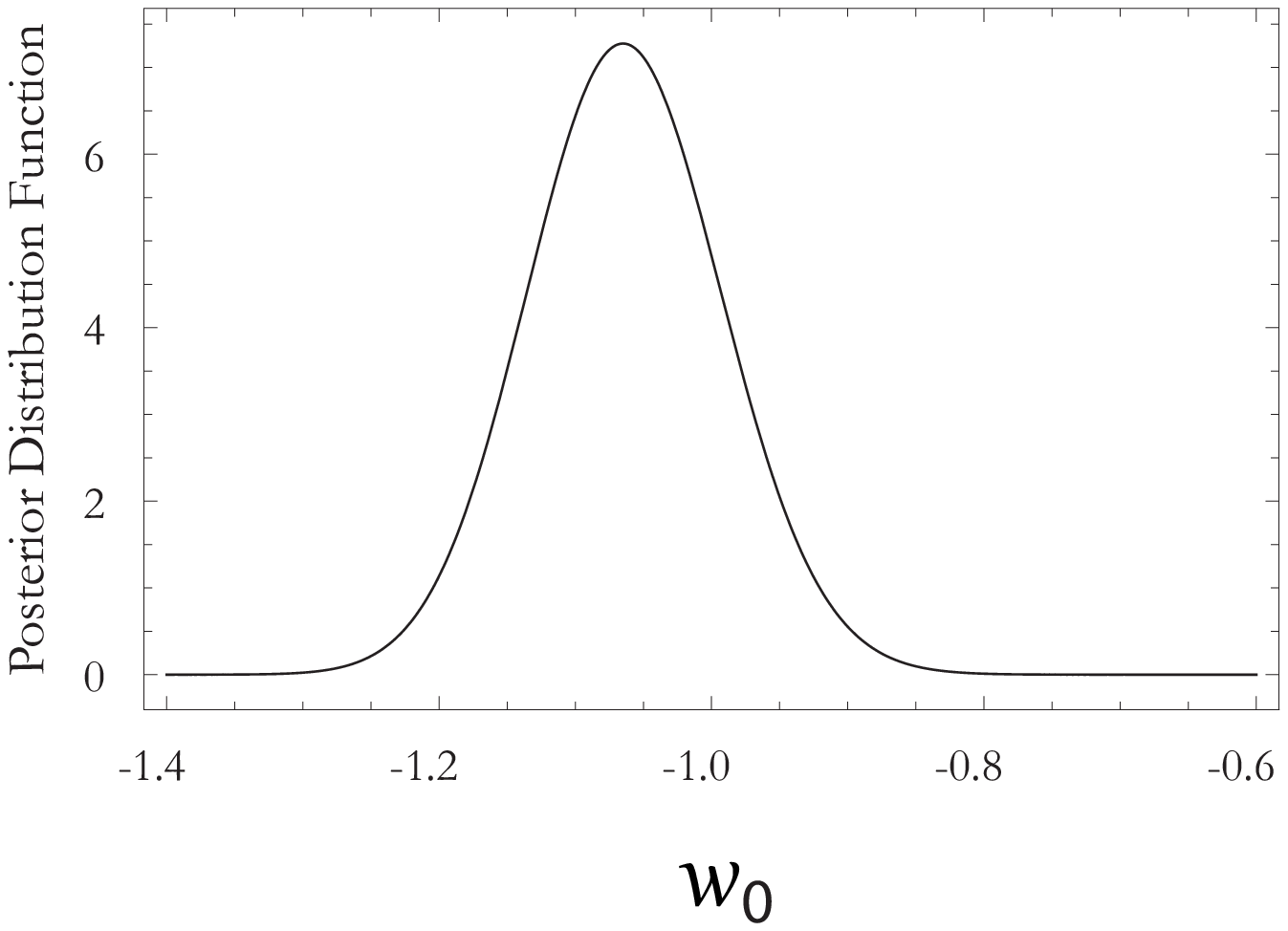}
  } % Termina de incluir a figura fig1.pdf
  \subfloat[]{ % ComeÃ§a a incluir a figura fig2.pdf na mesma linha da figura fig1.pdf
    \includegraphics[width=5.4cm,height=4.5cm]{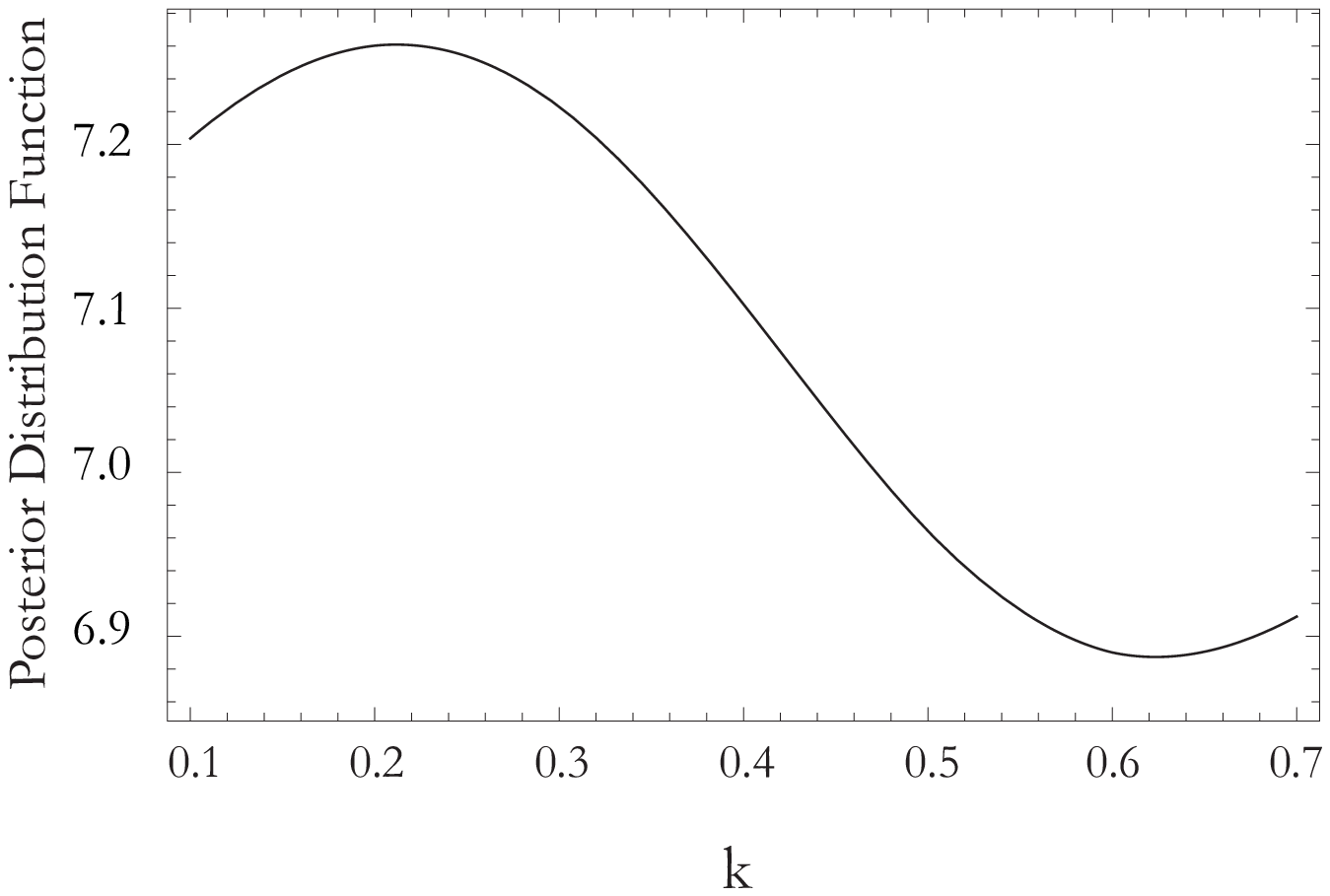} 
  } \\% Termina de incluir a figura fig2.pdf
  
  \subfloat[]{ % ComeÃ§a a incluir a figura fig2.pdf na mesma linha da figura fig1.pdf
    \includegraphics[width=5.4cm,height=4.5cm]{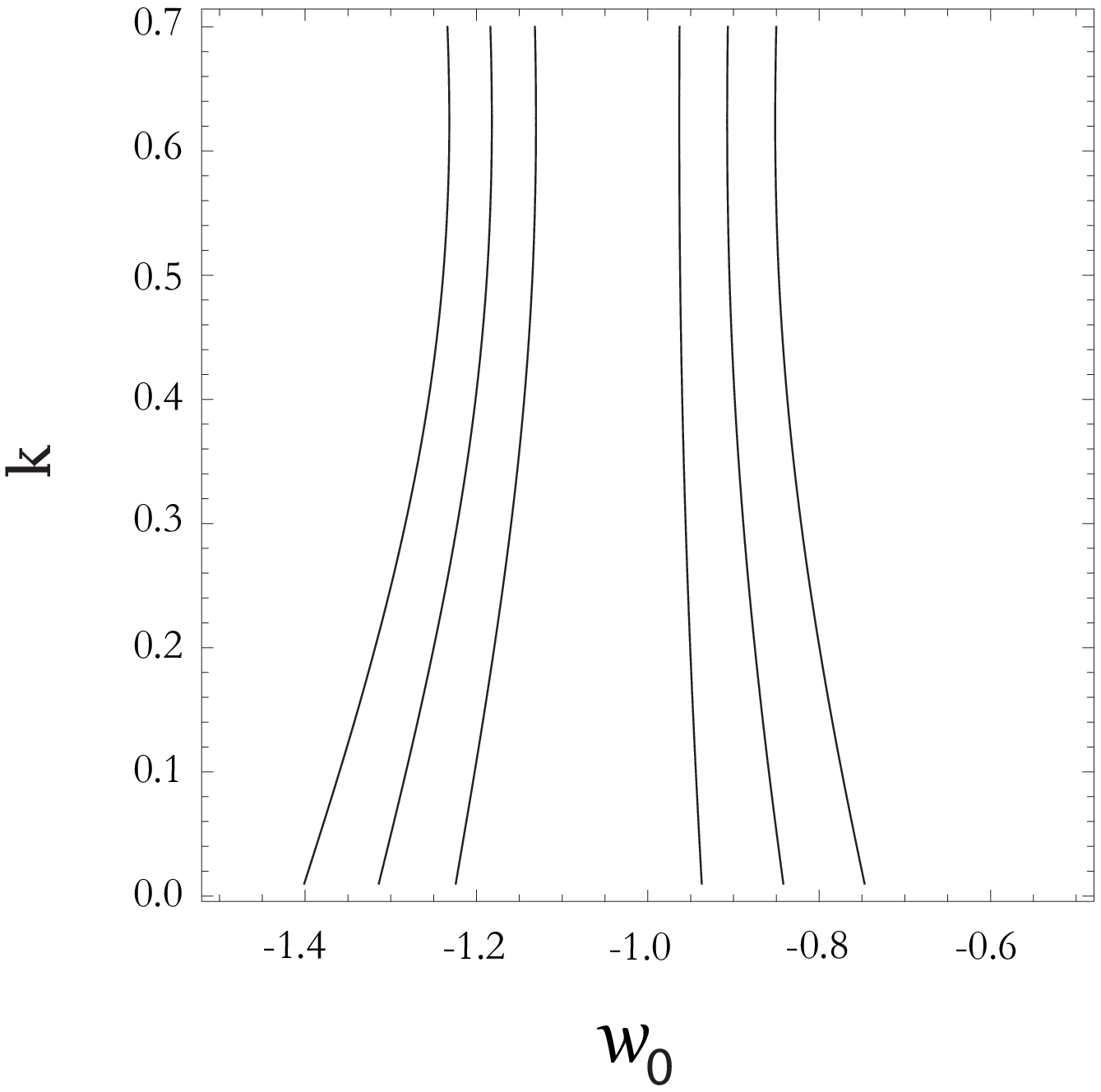}
  } % Termina de incluir a figura fig2.pdf
   % Com esse comando iremos incluir a Ãºltima figura na prÃ³xima linha
  \subfloat[]{ % ComeÃ§a a incluir a figura fig3.pdf na linha abaixo
    \includegraphics[width=5.4cm,height=4.5cm]{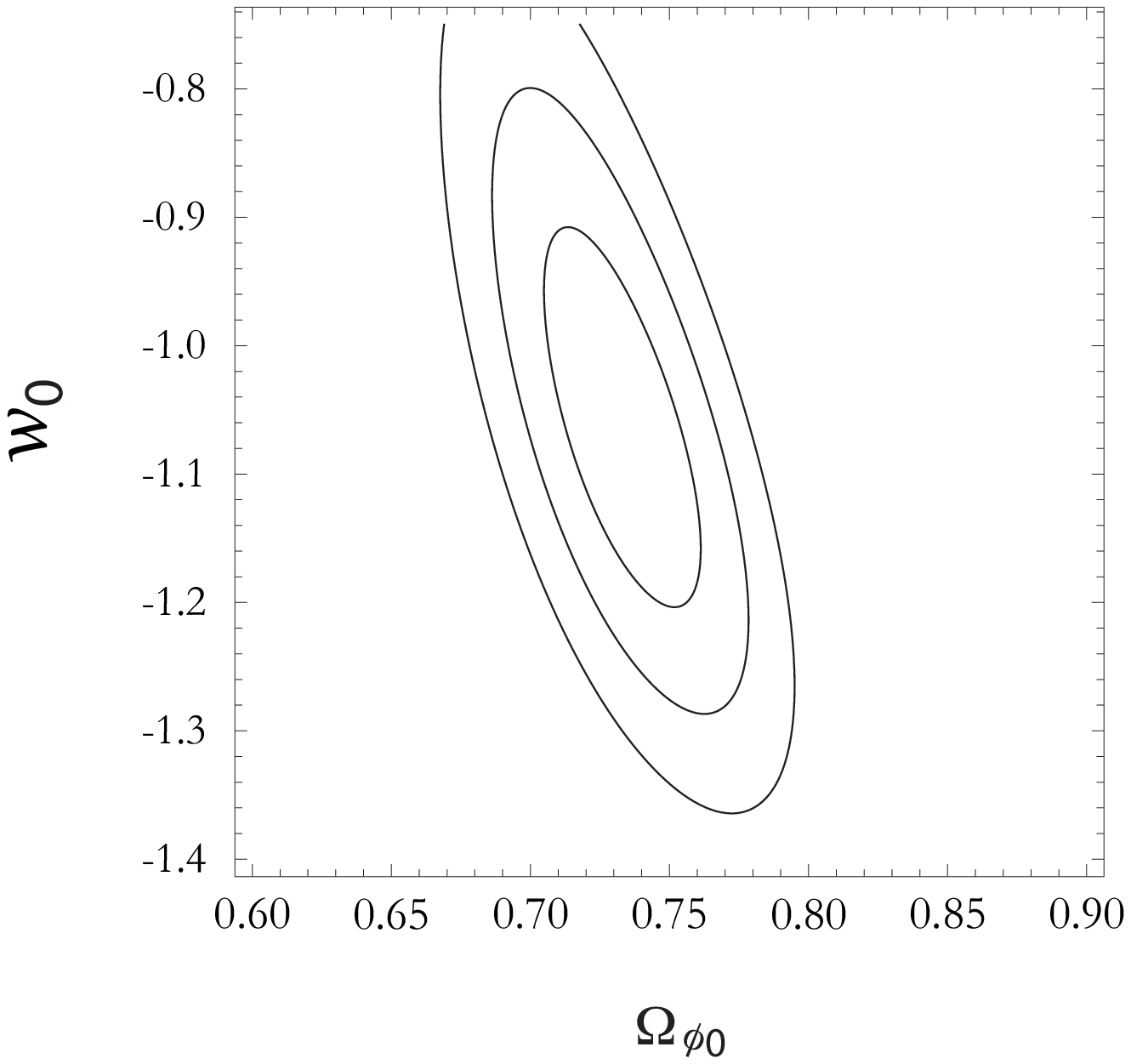}
  } 
 \\
  \caption{(a) PDF function for parameter $w_0$  after marginalization over $\Omega_{\phi 0}$ and $k$.
(b) PDF function for parameter $k$  after marginalization over $\Omega_{\phi 0}$ and $\omega_0$. (c) $w_0$ -$k$ plane and  
(d) $\Omega_{\phi 0}$ - $w_0$ plane. In (c) and (d) 
contour lines at 1$\sigma$, 2$\sigma$ and 3$\sigma$ 
CL are presented.}
\label{f7}
  \end{figure*}

Thus, considering Eqs. (\ref{eos}), (\ref{ka})  and
(\ref{pngb1})  we have performed a new analysis where free 
parameters are now $\Omega_{\phi 0}$, $w_0$ and $k$  where we have fixed $\phi_i/f =1.9$.  
Using SNIa alone our results give us constraints for $\Omega_{\phi 0}=0.76^{+0.06}_{-0.05}$, 
$w_0 = -1.16^{-0.14}_{+0.14}$ and $k=0.62^{+0.8}_{-0.62}$ ($2\sigma$ CL). 
By using $H(z)$ alone our results give us  constraints for $\Omega_{\phi 0}=0.72^{+0.05}_{-0.05}$, 
$w_0 = -0.83^{-0.30}_{+0.32}$ and $k=0.8^{+0.5}_{-0.8}$ ($2\sigma$ CL),
while the joint analyses gives us $\Omega_{\phi 0}=0.73^{+0.04}_{-0.04}$, $w_0 = -1.06^{+0.11}_{-0.11}$
and $k=0.25^{+0.9}_{-0.25}$ ($2\sigma$ CL) with  $\chi^2_\nu \sim 1.14$. 
Results of the joint analyses are presented in FIG.(\ref{f7}).  FIG. \ref{f7} (a)  
shows us the PDF function for parameter $w_0$  after marginalization over $\Omega_{\phi0}$ and $k$ 
FIG. \ref{f7} (b)  shows us the PDF function for parameter $k$ 
after marginalization over $\Omega_{\phi 0}$ and $w_0$ . 
FIG. \ref{f7} (c)  shows us the $w_{0}$ - $k$ plane with 
contour lines at 1$\sigma$, 2$\sigma$ and 3$\sigma$ 
confidence levels. FIG.\ref{f7} (d) visualizes the $\Omega_{\phi 0}$ - $w_0$  
plane with  contour lines at 1$\sigma$, 2$\sigma$ and 3$\sigma$ 
confidence levels. Note that there is a high degeneration in $k$-parameter. 

In order to see how our results are affected by $\phi_i/f$, we have varied such a parameter in the range
$0.6<\phi_i/f<2.5$ and we redone the analysis. Results indicate that  best-fit values for $\Omega_{\phi_0}$
and $w_0$ vary in $\sim 5\%$ as maximum. However, best-fit value for $k$ varies $\sim 40\%$. Such high 
variation is related to the strong degeneration in $k$.
 
 % Fecha o ambiente de figuras

%%%%%%%%%%%%%%%%%%%%%%%%%%%%%%%%%%%%%%%%%%%%%%%%%%%%%%%%5
\section{Conclusions}
\label{concl}
%%%%%%%%%%%%%%%%%%%%%%%%%%%%%%%%%%%%%%%%%%%%%%%%%%%%%%%
In this work an elliptic generalization of the 
classical pseudo Nambu--Goldstone
potential is studied. We have denominated 
such a generalization  the {\it Jacobian pseudo Nambu--Goldstone} 
(JpNGB) potential.  The JpNGB is used as an inflationary potential and we discuss the general conditions of the parameters to describe an inflationary stage. 
Additionally, we have calculated the values
to the inflaton necessary to obtain the number of e--folds of physical
expansion required to generate the seeds for the creation of the large 
structures in the iniverse. Also,  the relevant quantities for scalar and tensor 
perturbations, i.e., the scalar spectral index $n_s$ and the scalar--to--tensor ratio 
$r$, have been calculated explicitly and then contrasted with recent observational data 
from the {\it Planck collaboration}, where marginalized joint $68\%$ and $95\%$
confident level regions for Planck plus WMAP data for the model $\Lambda$CDM 
plus $r$ were used. These analyses predict that our generalization is able
to describe the inflationary stage  with exception of potentials with  
$k\sim 1$.

Afterward,  we have used the JpNGB potential and  the
construction of a family of quintessence models parametrized
by  parameter $k$ was implemented, where pNGB quintessence is the particular case $k=0$. 
We solved equations for generalized potential  in order to study the
evolution of the dynamical EoS and its evolution with the scale
factor. It was found that the same initial conditions lead us to a different 
evolution  of EoS for each different $k$,  opening the possibility 
of discriminating between them by using observational data. 

As  is well known, pGNB is a typical thawing quintessence model. 
In order to see if
similar behavior appears for the family of  generalized 
models, we performed a study of dynamical phase space $w'-w$.
It was confirmed that such a generalization maintains the thawing features 
of the  pNGB for all the possible values of $k$. This behavior allows 
us to use an analytical approximation of the dynamical $w(a)$ 
recently discovered for thawing models. 

In order to ascertain whether the generalization is compatible
with the observations, we performed an observational analyses using
SNIa and $H(z)$ data and the analytical solution for $w(a)$ (Eq. (\ref{eos})).
After marginalization over $h$, the space of parameters in this case 
is $(\Omega_{\phi 0},w_0,K)$. The best-fit values using SNIa and $H(z)$ 
joint analyses are  $\Omega_{\phi 0}=0.73^{+0.04}_{-0.05}$, $w_0 = -1.05^{+0.13}_{-0.17}$
and $K=0.99^{+8.2}_{-0.99}$ ($2\sigma$ CL) with a  $\chi^2_\nu \sim 1.1$. These results  are in agreement with previous analyses.
However, in general, the $K$-parameter is not adequate to 
discriminate between different thawing models  through the data.
Since  we have been  interested in knowing if some generalized  model is preferred or 
ruled out by the observations, it is necessary to consider $K$  as a function of $k$. 
Thus, after marginalization over $h$ and using a joint analyses with SNIa and $H(z)$, 
we have as best-fit values $\Omega_{\phi 0}=0.73^{+0.04}_{-0.04}$, $w_0 = -1.06^{-0.11}_{+0.11}$ 
and $k=0.25^{+0.9}_{-0.25}$ ($2\sigma$ CL) with a  $\chi^2_\nu \sim 1.14$.
These results show us that at 2$\sigma$ CL,  JpNGB models with  $0\leq k \leq1$ are not ruled out by 
SNIa and $H(z)$ data,i.e., quintessence fields with  potential (\ref{pngb1}) with the 
$\sim 70\%$ of the content of the dark energy  and with $w \sim -1$ are compatible 
with SNIa and $H(z)$ data.  Nevertheless  it seems to exist a strong degeneration 
in  the $k$-parameter which needs to be broken  using another data sets
and with a  detailed  study at perturbative level. 

Finally, we can conclude that  the generalization considered here is a viable model
and it  works well overall for the inflationary era and  for the acceleration era in late times.
Of course a better and a more detailed study is needed that considers the limits of approximation for (\ref{eos}) 
and exact numerical solutions as well as other data sets like CMB or matter power 
spectrum. Such points are still under investigation.

\section*{Acknowledgments}
The authors wish to thank V\'ictor H. C\'ardenas, Ram\'on Herrera and Winfried Zimdahl
for useful comments and discussions. J. R. V.  thanks the Universidade 
Federal do Esp\'irito Santo, while working on this paper. 
This research is also supported by Comisi\'on Nacional de Investigaci\'on Cient\'ifica 
y Tecnol\'ogica through FONDECYT grants No  11130695. WSHR was  
supported by Brazilian agencies FAPES (BPC No 476/2013) at the begining of 
this work,  and  CAPES at the end (proccess No 99999.007393/2014-08).
WSHR is grateful for the hospitality of the Physics Department of McGill University.

\appendix
\section{ A brief review of Jacobian elliptic functions}\label{app:jef}
The Jacobi elliptic functions are set of elliptic functions whose
three basic functions are called the {\it Jacobi elliptic cosine} 
cn\,$(u, k)$, {\it Jacobi elliptic delta}  dn\,$(u, k)$, and 
{\it Jacobi elliptic sine} sn\,$(u, k)$, where $k$ 
is known as the elliptic modulus. They arise from the inversion of the elliptic integral of the first kind \cite{byrd,hancock,Armitage,Mey0l,tablas}
\begin{eqnarray}\nonumber
u(y, k)\equiv u&=&\int_0^y\frac{dt}{\sqrt{(1-t^2)(1-k\,t^2)}}\\
&=&
\int_0^{\varphi}\frac{d\theta}{\sqrt{1-k\sin^2\theta}}=
F(\varphi, k),\label{a1}
\end{eqnarray}
which was
studied and solved by Abel and Jacobi.
The trigonometric and hyperbolic limit are
\begin{eqnarray}
&&\textrm{sn} (u, 0)=\sin u,\quad \textrm{sn} (u, 1)=\tanh u,\\
&&\textrm{cn} (u, 0)=\cos u,  \quad \textrm{cn} (u, 1)=\textrm{sech}\, u,\\
&& \textrm{dn} (u, 0)=1,\, \qquad \,\,\textrm{dn} (u, 1)=\textrm{sech}\, u.
\end{eqnarray}
The quotients and reciprocal of $\{\textrm{sn}\, u,\, \textrm{cn}\, u,\, \textrm{dn}\, u\}$
are designated in {\it Glaisher's notation} by
\begin{eqnarray}
&&\textrm{ns}\, u=\frac{1}{\textrm{sn} \, u},\quad \textrm{cs}\, u=\frac{\textrm{cn} \, u}{\textrm{sn} \, u},\quad \textrm{ds}\, u=\frac{\textrm{dn} \, u}{\textrm{sn} \, u},\\
&&\textrm{nc}\, u=\frac{1}{\textrm{cn} \, u},\quad \textrm{sc}\, u=\frac{\textrm{sn} \, u}{\textrm{cn} \, u},\quad \textrm{dc}\, u=\frac{\textrm{dn} \, u}{\textrm{cn} \, u},\\
&& \textrm{nd}\, u=\frac{1}{\textrm{dn} \, u},\quad \textrm{sd}\, u=\frac{\textrm{sn} \, u}{\textrm{dn} \, u},\quad \textrm{cd}\, u=\frac{\textrm{cn} \, u}{\textrm{dn} \, u}.
\end{eqnarray}
Therefore, in all, we have twelve Jacobian elliptic functions.
Finally, some useful fundamental relations between Jacobian elliptic functions
are 
\begin{eqnarray}
&&\textrm{sn}^2 u +  \textrm{cn}^2 u=1,\\
&&\textrm{dn}^2 u+ k\, \textrm{sn}^2 u  =1,\\
&& \textrm{dn}^2 u-k\,\textrm{cn}^2 u=k',\\
&& \textrm{cn}^2 u+k'\,\textrm{sn}^2 u =\textrm{dn}^2 u.
\end{eqnarray}

\end{document}